\documentclass[reprint, amsmath, amssymb, aps, showpacs, superscriptaddress]{revtex4-1}

\usepackage{float, graphicx}
\usepackage{dcolumn}
\usepackage{bm}
\usepackage{array}
\usepackage{tabularx,ragged2e,booktabs,caption}
\usepackage{hyperref}

\begin{document}


\title{Probing electron-argon scattering for liquid-argon based neutrino-oscillation program}
\thanks{Presented at the International Workshop on {\it{(e,e'p)}} Processes, July 2-6, 2017, Bled, Slovenia.}
\author{V.~Pandey}
\email{Speaker, email: vishvas@vt.edu}
\affiliation{Center for Neutrino Physics, Virginia Tech, Blacksburg, Virginia 24061, USA}
\author{D. Abrams}
\affiliation{Department of Physics, University of Virginia, Charlottesville, VA, 22904, USA}
\author{S.~Alsalmi}
\affiliation{Department of Physics, Kent State University, Kent, OH 44242, USA}
\author{A.~M.~Ankowski}
\affiliation{SLAC National Accelerator Laboratory, Stanford University, Menlo Park, CA, 94025, USA}
\author{J.~Bane}
\affiliation{Department of Physics and Astronomy, University of Tennessee, Knoxville, Tennessee 37996, USA}
\author{O.~Benhar}
\affiliation{INFN and Department of Physics, ``Sapienza'' Universit\`a di Roma, I-00185 Roma, Italy}
\author{H.~Dai}
\affiliation{Center for Neutrino Physics, Virginia Tech, Blacksburg, Virginia 24061, USA}
\author{D.~B.~Day}
\affiliation{Department of Physics, University of Virginia, Charlottesville, VA, 22904, USA}
\author{D.~W.~Higinbotham}
\affiliation{Thomas Jefferson National Accelerator Facility, Newport News, VA, 23606, USA}
\author{C.~Mariani}
\affiliation{Center for Neutrino Physics, Virginia Tech, Blacksburg, Virginia 24061, USA}
\author{M.~Murphy}
\affiliation{Center for Neutrino Physics, Virginia Tech, Blacksburg, Virginia 24061, USA}
\author{D. Nguyen}
\affiliation{Department of Physics, University of Virginia, Charlottesville, VA, 22904, USA}

\collaboration{for the Jefferson Lab E12-14-012 Collaboration}


\begin{abstract}
The electron scattering has been a vital tool to study the properties of the target nucleus for over five decades. Though, the particular interest on $^{40}$Ar nucleus stemmed from the progress in the accelerator-based neutrino-oscillation experiments. The complexity of nuclei comprising the detectors and their weak response turned out to be one of the major hurdles in the quest of achieving unprecedented precision in these experiments. The challenges are further magnified by the use of Liquid Argon Time Projection Chambers (LArTPCs) in the short- (SBN) and long-baseline (DUNE) neutrino program, with almost non-existence electron-argon scattering data and hence with no empirical basis to test and develop nuclear models for $^{40}$Ar. In light of these challenges, an electron-argon experiment, E12-14-012, was proposed at Jefferson Lab. The experiment has recently successfully completed collecting data for $(e,e'p)$ and $(e,e')$ processes, not just on $^{40}$Ar but also on $^{48}$Ti, and $^{12}$C targets. While the analysis is running with full steam, in this contribution, we present a brief overview of the experiment.
\end{abstract}


\maketitle


\section{\label{sec:introduction}Introduction}

Most of our current knowledge about the complexity of nuclear environment - nuclear structure, dynamics, and reaction mechanisms - has been accumulated by studying electron scattering off target nuclei. The electron scattering of nucleus, governed by quantum electrodynamics, has advantage over the proton or pion scattering off nuclei which are dominated by strong forces. The electromagnetic interaction is well known within quantum electrodynamics and is weak compared to hadronic interaction and hence the interaction between the incident electron and the nucleus can be treated within the Born approximation, i.e. within a single-photon exchange mechanism. 

In the last few decades, a wealth of high precision electron scattering data has been collected, over a variety of nuclei ranging from $^{3}$He to $^{208}$Pb, at several facilities including Bates, Saclay, Frascati, DESY, SLAC, NIKHEF, Jefferson Lab, etc., among others. The ability to vary electron energy, and scattering angle and hence the energy and moment transferred to the nucleus ($\omega, q$) - combined with the advancement in high-intensity electron beams, high performance spectrometers and detectors - resulted into investigating processes ranging from quasi-elastic (QE) to the $\Delta$ resonance to complete inelastic (resonant, non-resonant, and the deep inelastic scatterings (DIS)) with significant details. A number of those datasets were further utilized to separate the longitudinal and transverse response functions through the Rosenbluth separation. Several decades of experimental work has provided sufficient testbed to assess and validate several theoretical approximations and predictions and hence propelled the theoretical progress staged around nuclear ground state properties, nuclear many-body theories, nuclear correlations, form factors, nucleon-nucleon interactions, etc. For a detailed recent review on electron scattering, we refer readers to Ref.~\cite{Benhar:Review-2008}, a web-archive of accumulated data is maintained at Ref.~\cite{Day:Archive}.

The electron scattering, besides being immensely interesting in itself, turned out to be of great importance for accelerator-based neutrino-oscillation experiments. The quest of measuring neutrino oscillation parameters with an unprecedented level of precision, search for CP violation in leptonic sector, and exploration of a possible fourth generation of neutrinos - relies greatly on an accurate description of the nucleus and its electroweak response. To this end, the data collected with electron-nucleus scattering has provided the benchmark to test the nuclear models that can be further extended to neutrino-nucleus scattering. The extension of the formalism from electron-nucleus scattering, where only vector current contributes, to neutrino is rather straightforward with the main difference being the additional axial current contribution to neutrino scattering. Despite the fact that (unpolarized) electron scattering provide access to only vector response, the vector current is conserved between electromagnetic and weak response through conserved vector current (CVC). 

The current motivation for another electron scattering experiment, especially on $^{40}$Ar nucleus, emerged from the surge in the progress and need of the liquid-argon time projection chamber (LArTPC) based short- (SBN)~\cite{SBNProposal:2015} and long-baseline (DUNE)~\cite{DUNEReport:2015} neutrino program. The current status of systematic uncertainties even with the relatively well-know isospin symmetric nuclei such as $^{12}$C and $^{16}$O (pertinent for water Cherenkov and mineral-oil detectors) of which a range of electron scattering data is available, are already of the order of 10\%. The magnitude of uncertainty is expected to rise significantly considering almost non-existence (an inclusive cross section measurement performed at Frascati is the only data available on argon~\cite{Anghinolfi:1995}) electron scattering studies on argon, and hence no empirical testbed available to develop accurate models. Furthermore, in comparison to otherwise widely used isospin symmetric nuclei, argon consists of few extra neutrons than protons which may impact the test for the CP violation if neutrino and antineutrino behold different nuclear effects on argon.

In light of these needs, a dedicated electron experiment on $^{40}$Ar, E12-14-012, is recently proposed at Jefferson lab~\cite{Proposal:2014}. We aim to measure
spectral functions (SF) of $^{40}$Ar, as well as to measure the inclusive and exclusive cross sections on $^{40}$Ar, $^{48}$Ti, and $^{12}$C. The SF can provide indispensable information about the initial energy and momentum of the nucleons bound in the argon nucleus which can be utilized in the reconstruction of neutrino energy
and accurate predictions of event rates. In addition, measuring SF of argon nucleus will provide a vital input in the further development of a unified nuclear model based on nuclear many-body theory and the spectral function formalism~\cite{Benhar:2005}, that has been successful in describing the electron scattering in a variety of kinematics~\cite{Ankowski:2015, Rocco:2016, Vagnoni:2017}. Nevertheless, a new high precision cross section data will provide a bench mark to test and validate the theoretical models that have recently reported neutrino scattering on argon nucleus~\cite{VanDessel:2017, Pandey:2014, Butkevich:2012} or the ones intended to extend their models to neutrino-argon scatterings. 

The article is structured as follows. In Sec.~\ref{sec:uncertainities}, a brief discussion on the impact of nuclear effects on the accelerator-based neutrino-oscillation experiments is presented. We will then move on to give an overview of the E12-14-012 experiment in Sec.~\ref{sec:e12-14-012}. The summary is presented in Sec.~\ref{sec:summary}.


\section{\label{sec:uncertainities}The Impact of nuclear effects on neutrino-oscillation experiments}

The systematic challenges in neutrino experiments are manifold. The energy of the interacting neutrino is not known, the kinematics of the interaction in the target nucleus is not known and the only known (to some degree) quantity - the topology of the final state particles and their energy - is subjected to detector type, detection thresholds, and the accuracy of particle identification and background reduction processes. The necessity of an accurate nuclear model is essential at almost every step of the analysis. For recent reviews on this subject, we refer readers to Refs.~\cite{Benhar:Review-2015, Alvarez-Ruso:2017, Ankowski:2016}.

In the event of a neutrino oscillating from $\nu_i$ to $\nu_j$ and for a given observable topology,  the observed event rate at far detector is a convolution of neutrino flux at near detector ($\phi_{\nu_{i}}$), probability of oscillation from flavor $i$ to $j$, and the 
neutrino-nucleus cross section $\sigma_{\nu_j}$, and detection efficiency $\epsilon_{\nu_j}$ at far detector
\begin{equation}
 \mathcal{R}(\nu_i \rightarrow \nu_j) \propto \phi_{\nu_i} \otimes P(\nu_i \rightarrow \nu_j) \otimes \sigma_{\nu_j} \otimes \epsilon_{\nu_j} \label{rate}
\end{equation}
with oscillation probability, considering simple example of two neutrino flavors, given as: 
\begin{equation}
 P(\nu_i \rightarrow \nu_j) \simeq \sin^{2}2\theta \sin^{2}\left(\frac{\Delta m^{2} L}{4E}\right),
\end{equation}
with $\theta$ and $\Delta m^{2}$ are mixing angle, and the squared-mass difference, respectively. The accuracy of the measurement of the (energy-dependent) neutrino oscillation probability relies strongly on the accuracy with which a nuclear model (implemented in a Monte-Carlo event generator), can describe all neutrino-nucleus interaction types that can produce the observed topology (that depends both on initial and final state nuclear effects). The Monte Carlo simulations of neutrino interactions used in data analysis are typically based on relativistic Fermi gas (RFG) model, which cannot reproduce the relevant cross sections with sufficient accuracy. As it stands currently, a lack of reliable nuclear recipe contributes to the main source of systematic uncertainty and is considered one of the main hurdles in further increasing the obtained precision. 

Let us consider, for example, the case of neutrino energy determination. This situation differs dramatically from the electron scattering experiments, where the interacting charged lepton energy is precisely known on event-by-event basis. In neutrino scattering the neutrino energy can only be estimated based on what is observed in the detector. The observed event topology and energy, i.e. particle identification and their momentum are utilized, using a nuclear model, to identify the neutrino energy at the interaction vertex. Now, assuming a single-nucleon knockout process, the reconstructed neutrino energy can be expressed as
\begin{equation}
E_{\nu_{l}} =  \frac{M_{n}^{2} - m_{l}^{2} - E_{n}^{2} + 2 E_n E_l - 2 {{\vec p}_n}\cdot{{\vec k}_l} + |{\vec p}_n|^2} {2(E_n - E_l - |{\vec p}_n| \cos\theta_n + |{\vec k}_l| \cos\theta_l)}.
\end{equation}
Where, $m_l$, $E_l$, ${\vec k}_l$, and $\theta_l$ are the mass, energy, momentum, and scattering angle of the outgoing charged lepton, these are either known or measured quantities. $M_n$ is the mass of the nucleon, while ${\vec p}_n$ and $E_n$ are the momentum and energy of the interacting neutron. Existing simulation codes routinely approximate the energy and momentum of the nuclear system as: $p_n = 0$ , $E_n = M_n-\epsilon$, where $\epsilon$ (separation energy) is approximated as a constant number. In a realistic picture of a nucleus the struck nucleon carries a momentum and energy distribution which are characteristic of a specific target nucleus. These, for example, can be derived from the spectral functions of the nucleons and can significantly improve the accuracy with which the energy is reconstructed.


\begin{figure} [H]
\centering
\includegraphics[width=0.80\columnwidth]{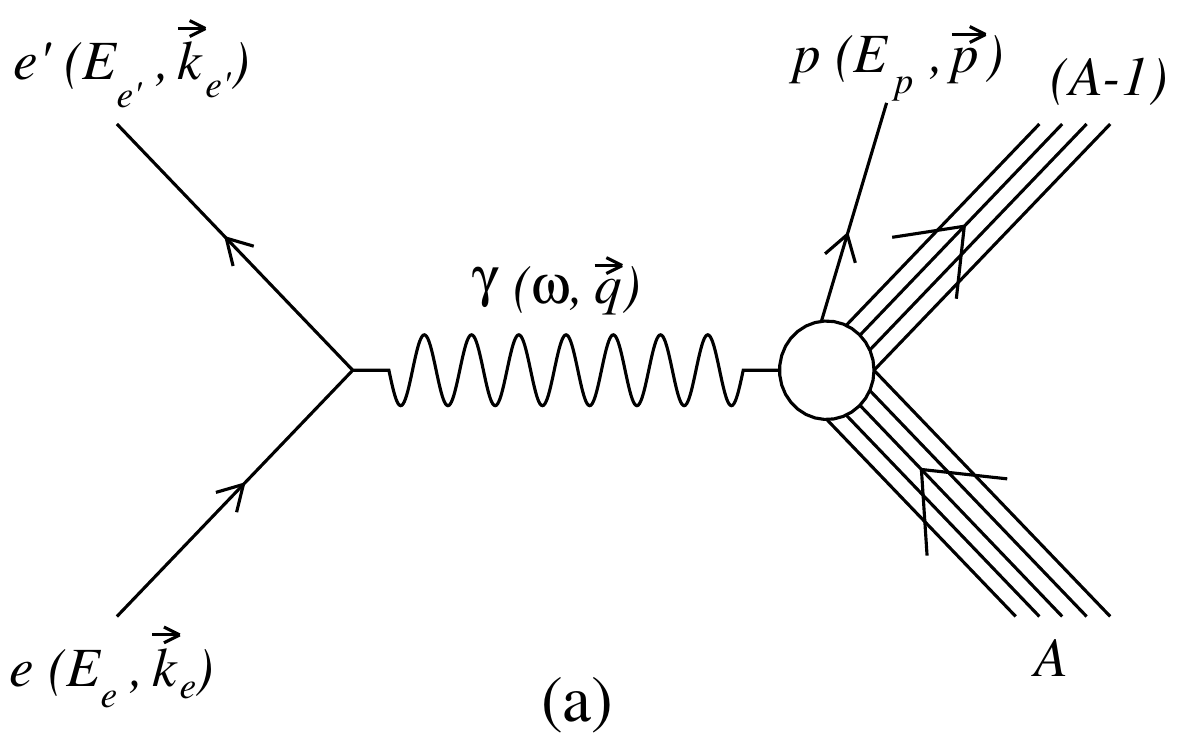}
\includegraphics[width=0.80\columnwidth]{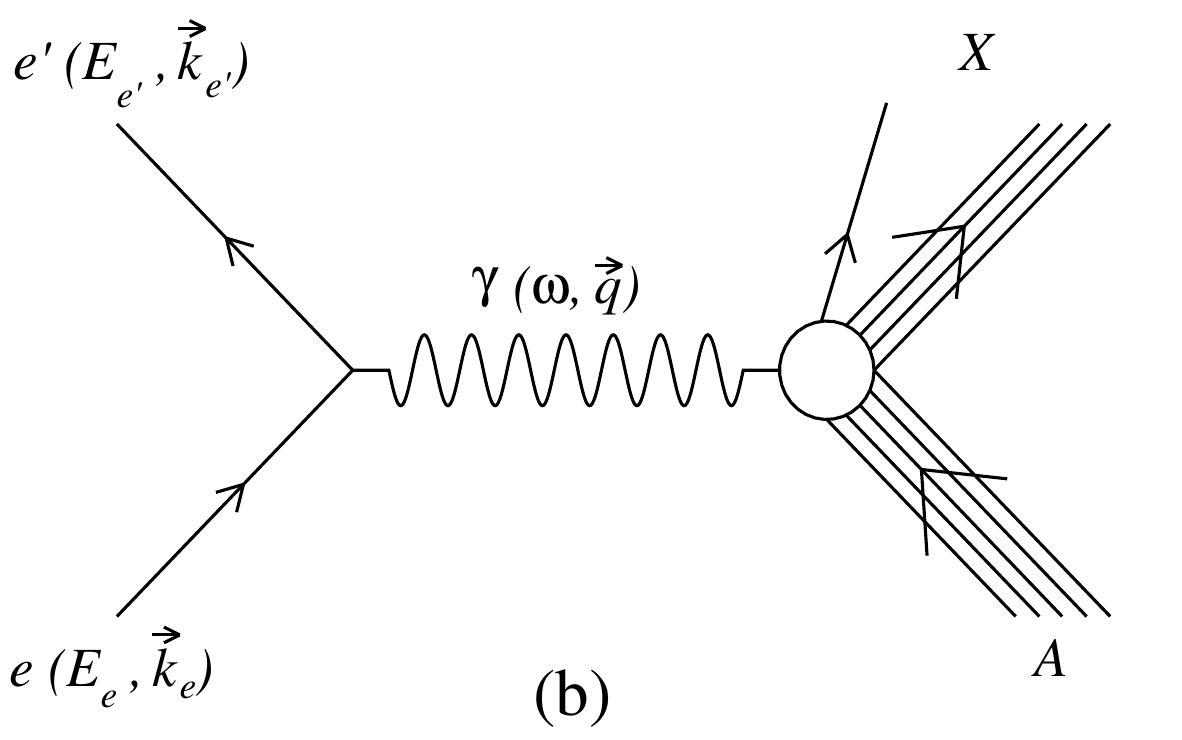}
\caption{Processes considered in the E12-14-012 experiment: (a) Coincidence $(e,e'p)$ reaction, where scattered electron and knocked out proton are measured in coincidence and the residual nucleus is left undetected, and (b) $(e,e')$ reaction, where only scattered electron is detected.}
\label{feynman_diagram}
\end{figure}

\section{\label{sec:e12-14-012}E12-14-012 experiment}
The E12-14-012 experiment was proposed in 2014~\cite{Proposal:2014}. The experiment was performed in Hall A of Jefferson Lab from February to March 2017. We initially proposed measurements on a coincidence $(e,e'p)$ scattering off $^{40}$Ar. Though, the physics reach is later extended to additional target nucleus, $^{48}$Ti and $^{12}$C, and included an inclusive measurement for a single kinematics.

\subsection{$(e,e'p)$ and $(e,e')$ Reactions} 
 
We considered coincidence $(e,e'p)$ process, as shown in Fig~\ref{feynman_diagram}(a), where an electron of four-momentum $k_e \equiv (E_e, {\vec k}_e)$ scatters off a nuclear target $A$ and knocks-out a proton of four-momentum $p \equiv (E_p, {\vec p})$, leaving the residual $(A-1)$ system in any (undetected) bound or continuum state: 
\begin{equation}
 e + A \rightarrow e' + p + (A-1).
\end{equation}
The knocked out proton is detected in coincidence with the scattered electron of four-momentum $k_{e'} \equiv (E_{e'}, {\vec k}_{e'})$. The transferred four-momentum of the process is given as
\begin{equation}
q = k_e - k_{e'} \equiv (\omega, {\vec q})
\end{equation}

Such a reaction is a valuable source of information on the single-particle aspects of nuclear structure where one can define missing energy, $E_m$, from the energy conservation. Now, within the Plane Wave Impulse Approximation (PWIA) scheme, i.e. on the assumptions that at momentum transfers $q \gg \frac{1}{d}$ where d is the average inter-nucleon distance - the scattering from the nucleus can be reduced to incoherent sum of scattering from individual nucleons and that the final state interaction (FSI) between the produced hadron and the residual nucleus are negligible, the missing momentum, ${\vec p}_m$, can be estimate as
\begin{equation}
 {\vec p}_m = {\vec p} - {\vec q} 
\end{equation}
and the missing energy, in the limit of low missing momentum, can be reduced to
\begin{equation}
 E_m = {\omega} - T_p 
\end{equation}
where $T_p = E_p - m$, is the kinetic energy of the outgoing proton. It is worth pointing out that despite PWIA scheme can provide a clear manifest of the description of the $(e,e'p)$ reaction, the corrections arising from the FSI between outgoing nucleon and the residual nucleus are not negligible. These, for example, can be treated within the Distorted Wave Impulse Approximation (DWIA) description~\cite{Giusti:1993}, and will be taken care of in the analysis of the $(e,e'p)$ data.

The total Spectral Function (SF), which describes the probability distribution of finding a nucleon with momentum $p_m$ and removal energy $E_m$ in the target ground state, can be expressed as the combination of experimentally measured mean-field (MF) and theoretically calculated correlation part
\begin{equation}
 P(p_m, E_m) = P_{MF} (p_m, E_m) + P_{corr} (p_m, E_m).
\end{equation}
The mean-field part of the SF, $P_{MF} (p_m, E_m)$, which corresponds to low missing momentum and energy region and hence where the shell model dynamics dominates, can be extracted from coincidence $(e,e'p)$ data. While correlation part of the SF, $P_{corr} (p_m, E_m)$, which corresponds to the ground state nucleon-nucleon (short-range) correlations, is evaluated from theoretical calculations of uniform nuclear matter and Local Density Approximation (LDA).

The $(e,e')$ process, as shown in Fig~\ref{feynman_diagram}(b), where an electron of four-momentum $k_e \equiv (E_e, {\vec k}_e)$ scatters off a nuclear target $A$
 \begin{equation}
 e + A \rightarrow e' + X
\end{equation}
and only scattered electron is detected. Here, the cross section includes all available final states.

\begin{figure} [H]
\centering
\includegraphics[width=0.99\columnwidth]{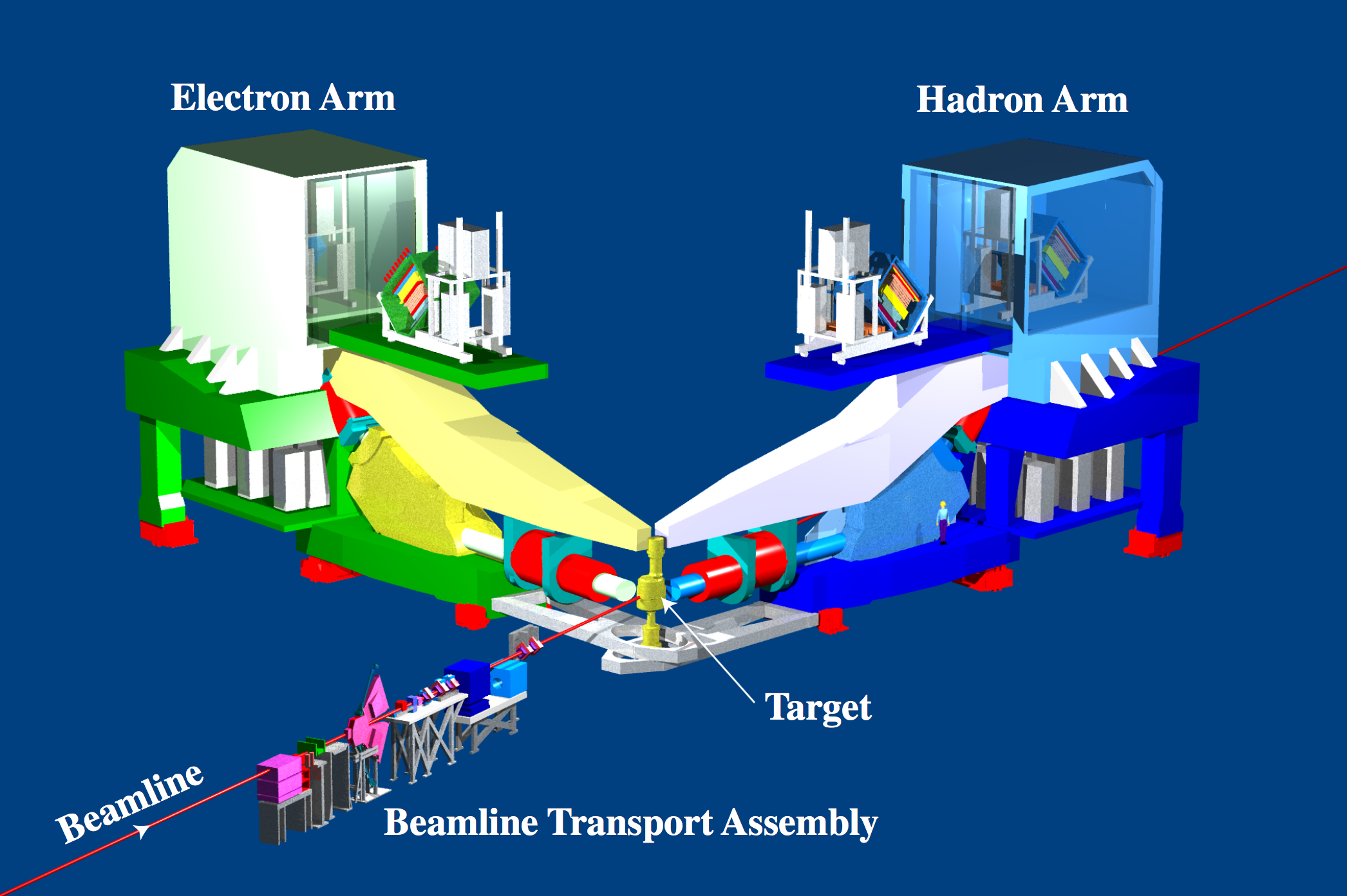}
\caption{A schematic layout of Hall A: representing the beamline, and both the Left and the Right High-Resolution Spectrometers as electron and hadron arms, respectively.}
\label{HallA}
\end{figure}

\subsection{Experimental Setup} 
 
A schematic layout of Hall A is shown in Fig.~\ref{HallA}. The electron beam of energy $2.222$ GeV is provided by the Continuous Electron Beam Accelerator Facility (CEBAF) at JLab. There are multiple diagnostic tools along the beamline such as beam current and position monitors. The beam current is monitored using two Beam Current Monitors (BCMs) which are resonant RF cavities where one cavity is denoted as upstream and other as downstream based on their relative positions along the beamline. The position of the beam, in the plane transverse to the beam direction at the target, is monitored by two Beam Position Monitors (BPMs). Beam position determination is important for vertex reconstruction and momentum calculation of scattered electron. 

The scattered electron goes to the Left High-Resolution Spectrometer (LHRS) and the scattered hadron goes to the Right High-Resolution Spectrometer (RHRS). HRS has the capability of running at high luminosity, and has a high resolution in both the momentum and the angle reconstruction of the scattered particles. The scattered particles go through two superconducting quadrupole magnets (which focus the charged particles), followed by a superconducting dipole magnet (that bends the charged particles by 45 degrees), the particle further passes through another quadrupole magnet. Superconducting magnets allow a large acceptance in both angle and momentum and provide a good resolution in position and angle. 

After passing through the magnet configuration, the particle enters the detector stack consists of a number of sub packages. There are scintillator planes which are used to form the trigger to activate the data-acquisition system and for the precise timing information of time-of-flight measurements and coincidence determination. Vertical Drift Chambers are used for particle tracking, i.e. to determine the position and direction of the scattered particles. The particles are identified using a variety of Cherenkov type detectors (aerogel and gas) and lead-glass shower counters. For a detailed description of Hall A setup, we refer readers to Ref.~\cite{Alcorn:2004}.

\begin{figure} [H]
\centering
\includegraphics[width=0.40\columnwidth]{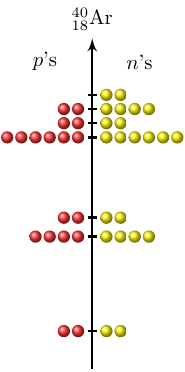}
\includegraphics[width=0.40\columnwidth]{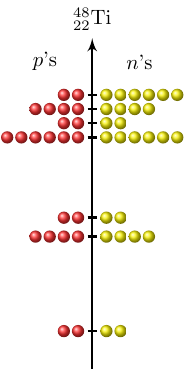}
\caption{Shell model structure of $^{40}$Ar and $^{48}$Ti nucleus. Note that the proton structure of titanium mirrors to that of the neutron structure of argon.}
\label{argon_titanium}
\end{figure}

\begin{table} [H]
\begin{center}
\begin{tabular}{|c|c|c|c|c|c|}  
\hline
         & $E_{e}$ & $E_{e'}$  & $\theta_{e}$  & $P_{p}$  & $\theta_{p}$ \\
         &  (MeV) & (MeV) & (deg)  & (MeV/c)  & (deg) \\  
\hline 
 \bf{Exclusive}  \\
\hline
 kin 1 & 2222 & 1799 & 21.5  & 915  & -50.0   \\  
\hline
 kin 2 & 2222 & 1716 & 20.0  & 1030  & -44.0  \\  
\hline
 kin 3 & 2222 & 1799 & 17.5  & 915  & -47.0   \\  
\hline
 kin 4 & 2222 & 1799 & 15.5  & 915  & -44.5  \\  
\hline
 kin 5 & 2222 & 1716 & 15.5  & 1030  & -39.0  \\  
\hline
 \bf{Inclusive} \\   
\hline
 kin 5 & 2222 &  - & 15.5  &  -  &  -  \\ 
\hline
\end{tabular}
\caption{The exclusive and inclusive kinematics for which the electron scattering data is collected on different target nucleus.}
\end{center}
\end{table}

\subsection{Nuclear Targets and Kinematic Setup} 

The proton spectral function of argon nucleus can be extracted from coincidence $(e,e'p)$ reaction on $^{40}$Ar. From the proton SF, the energy and momentum distribution of protons in the argon can be achieved and can be used in the reconstruction of antineutrino energy in liquid argon. For the reconstruction of neutrino energy in liquid argon detectors, one requires the neutron spectral function of argon. To this end, we introduced the idea of titanium nucleus, as can be seen in Fig.~\ref{argon_titanium}, the proton level structure of titanium mirrors to that of the neutron level structure of argon. Utilizing this level correspondence, the neutron SF of argon can be obtained from the proton SF extracted from $^{48}$Ti$(e,e'p)$. For the calibration of spectrometer pointing, the data is also collected on the single and multi foil optics target, $^{12}$C. Further data-sets are collected on the aluminum dummy target which will allow us to subtract the background generated at the end-caps of the target cell. 

The measurements are successfully performed at five different kinematics for exclusive reactions and one kinematics for inclusive reaction, as shown in Table~I. The incoming electron beam energy, $E_{e}$, the scattered electron energy, $E_{e'}$, (and therefore the energy transfer) and the momentum of knocked out proton, $P_{p}$, are held fixed. Now, by varying the electron scattering angle, $\theta_{e}$, (and moving the HRS accordingly), we spanned over different missing momentum. The kinematics are chosen to scan the region where shell model dynamics is known to dominate, and single-particle aspects of the nuclear structure can be probed. Note that the SF, that we aim to extract in this experiment, describes the ground state properties of the nucleus. Hence, it can be used in the description of any interaction involving single nucleon, independent of the final state. 


\section{\label{sec:summary}Summary}

The systematic uncertainties related to neutrino-nucleus signal in the detector present the biggest hurdle in achieving the precision goals of accelerator-based neutrino-oscillation experiments. To this end, the electron scattering data has been playing a significant role by providing the testbed to assess and validate different nuclear models intended to be used in neutrino experiments. Though, the challenges are appearing to amplify with the use of LArTPC in short- and long-baseline neutrino program with almost non-existence electron scattering data on argon nucleus, and hence ceasing any development towards an accurate neutrino-argon modeling. In light of these challenges, an electron argon experiment, E12-14-012, has recently been proposed at Jefferson lab. The experiment has recently been completed collecting data for $(e,e'p)$ and $(e,e')$ processes on $^{40}$Ar, $^{48}$Ti, and $^{12}$C targets for various kinematics. We hope that the Spectral functions and the cross sections extracted from this (only available high-precision) data on $^{40}$Ar nucleus will help reducing the systematic uncertainties, and will enable us achieving high-precision goals in the liquid-argon based short- and long-baseline neutrino program.


\acknowledgments

VP is grateful to the organizers of the workshop for the invitation and hospitality. VP, and CM acknowledge the support from the National Science Foundation under grant no. PHY$-$1352106. This work would not have been possible without the immense support of Hall A collaboration members of JLab, and without the contribution of the rest of the E12-14-012 collaboration members.




\begin{thebibliography}{}
\bibitem{Benhar:Review-2008}O.~Benhar, D.~Day, I.~Sick, Rev.~Mod.~Phys.~{\bf 80}, 189 (2008).
\bibitem{Day:Archive}O.~Benhar, D.~Day, and I.~Sick, arXiv:0603032 [nucl-ex]. \url{http://faculty.virginia.edu/qes-archive/index.html}
\bibitem{SBNProposal:2015}M.~Antonello {\it et al.} (MicroBooNE, LAr1-ND, and ICARUS-WA104 Collaborations), arXiv:1503.01520 [physics.ins-det].
\bibitem{DUNEReport:2015}R.~Acciari {\it et al.} (DUNE Collaboration), arXiv:1512.06148 [physics.ins-det].
\bibitem{Anghinolfi:1995}M.~Anghinolfi {\it et al.}, J.~Phys.~G{\bf 21}, L9 (1995).
\bibitem{Proposal:2014}O.~Benhar, {\it et al.} (Jefferson Lab E12-14-012 Collaboration), arXiv:1406.4080 [nucl-ex].
\bibitem{Benhar:2005}O.~Benhar, N.~Farina, H.~Nakamura, M.~Sakuda and R.~Seki, Phys.~Rev.~D{\bf 72}, 053005 (2005).
\bibitem{Ankowski:2015}A.~M.~Ankowski, O.~Benhar, M.~Sakuda, Phys.~Rev.~D{\bf 91}, 054616 (2015).
\bibitem{Rocco:2016}N.~Rocco, A.~Lovato, O.~Benhar, Phys.~Rev.~Lett.~{\bf 116}, 192501 (2016).
\bibitem{Vagnoni:2017}E.~Vagnoni, O.~Benhar, and D.~Meloni, Phys.~Rev.~Lett.~{\bf 118}, 142502 (2017).
\bibitem{VanDessel:2017}N.~Van Dessel, N.~Jachowicz, R.~Gonz\'{a}lez-Jim\'{e}nez, V.~Pandey and T.~Van Cuyck, arXiv:1704.07817 [nucl-th].
\bibitem{Pandey:2014}V.~Pandey, N.~Jachowicz, T.~Van Cuyck, J.~Ryckebusch and M.~Martini, Phys.~Rev.~C{\bf 92}, no. 2, 024606 (2015).
\bibitem{Butkevich:2012}A.~V.~Butkevich, Phys.~Rev.~C{\bf 85}, 065501 (2012).
\bibitem{Benhar:Review-2015}O.~Benhar, P.~Huber, C.~Mariani and D.~Meloni, Phys.~Rept.~{\bf 700}, 1 (2017).
\bibitem{Alvarez-Ruso:2017} L.~Alvarez-Ruso {\it et al.}, arXiv:1706.03621 [hep-ph].
\bibitem{Ankowski:2016}A.~M.~Ankowski and C.~Mariani, J.~Phys.~G{\bf 44}, no. 5, 054001 (2017).
\bibitem{Giusti:1993}S.~Boffi, C.~Giusti, and F.~D.~Pacati, Phys.~Rep.~{\bf 226}, 1 (1993).
\bibitem{Alcorn:2004}J.~Alcorn {\it et al.},  Nucl.~Instrum.~Meth.~A{\bf 522}, 294 (2004).
\end{thebibliography}
\end{document}